# A Multiscale Framework for In Silico Thrombus Generation and Photoacoustic Simulations

Hamed Ghodsi [1], Sara Cardona [2], Behrooz Fereidoonnezhad [2,*] and Sophinese Iskander-Rizk [1,*]

[1] Department of Precision and Microsystems Engineering, Faculty of Mechanical Engineering, Delft University of Technology, the Netherlands
[2] Department of BioMechanical Engineering, Faculty of Mechanical Engineering, Delft University of Technology, the Netherlands

*Senior authors contributed equally. Correspondence:

E-mail: b.fereidoonnezhad@tudelft.nl, S.Iskander-Rizk@tudelft.nl

## Abstract

Thrombus composition and microstructure play a critical role in determining the treatment success for thrombus-related diseases such as ischemic stroke and deep vein thrombosis. However, no in vivo diagnostic method can fully capture thrombus microstructure yet, hindering personalized treatment. Photoacoustic imaging is uniquely positioned to provide information on thrombi composition as it relays optical absorption information from diffuse photons at acoustic propagation depths. Computational modeling enables systematic exploration of microstructural effects on imaging signals, offering insights into developing improved in vivo diagnostic techniques. However, no photoacoustic simulation platform can model microstructural features within centimeter-scale phantoms at reasonable computational cost. In this work, we present REFINE, a topology-driven framework for generating in silico thrombi replicating its key replicating their key microstructural traits. Unlike existing methods, REFINE enables controlled, recursive optimization of thrombus topology, making it suitable for accurate photoacoustic modeling and potentially powerful for biomechanical analyses beyond this study. These digital thrombi are embedded into a multiscale photoacoustic simulation platform that bridges microscale acoustic modeling with macroscale thrombus geometries, enabling efficient and realistic simulation of photoacoustic signal responses. We successfully created unique representation of thrombi microstructure for various compositions and porosities. Our simulation framework effectively links microstructural features to macroscale imaging outcomes, in agreement with previous empirical studies. Our simulation results demonstrate that thrombus microstructure significantly affects photoacoustic spectral responses and can be reliably modeled in silico. These findings highlight the potential of a multiscale photoacoustic simulation approach as a powerful tool for characterizing tissue microstructure and demonstrate the utility of our computational framework for in silico thrombi analysis and the development of diagnostic imaging strategies.

# 1. Introduction

Thrombus-related diseases such as ischemic stroke, deep vein thrombosis, pulmonary embolism, and myocardial infarction are major public health concerns [1,2]. While treatment options like thrombolytics therapy and mechanical thrombectomy exist, selecting the most effective approach remains challenging [1,3–6]. One main obstacle is the wide variability in thrombus composition and microstructure across patients [7,8]. In a thrombectomy procedure, foreknowledge of thrombus characteristics and biomechanical properties is essential for effective treatment and for mitigating further complications [9–11]. Histological studies reveal that thrombi typically consist of red blood cells (RBCs) embedded within a fibrin mesh held together by platelets. However, their composition, spatial distribution of components, and biomechanical properties can vary significantly [8]. Experimental studies demonstrate a correlation between these microstructural properties of thrombi and their mechanical properties such as stiffness and rupture resistance [12–15]. Therefore, a better understanding of microstructural differences could help guide precise therapies and improve patient outcomes. However, there is no in vivo imaging method that can capture this level of detail yet, forcing clinicians to rely on limited data before procedures [16,17].

Photoacoustic imaging is a promising tool for visualizing compositional details in biological samples [18–21]. Moreover, preclinical studies demonstrate that photoacoustic imaging can detect microstructural differences in a variety of tissues [18,22–24]. The photoacoustic signal is closely tied to the sample's microstructure, even though issues such as light scattering, limited-view acquisition, and high-frequency ultrasound attenuation still pose challenges. This makes photoacoustic imaging a potential method for capturing thrombus microstructural variations. Additionally, minimally invasive thrombectomy procedures provide an opportunity to reach the thrombus using an intravascular photoacoustic probe, further positioning it to be a clinically viable solution [25,26].

Systematic experimental characterization of thrombus remains difficult due to limited sample availability and the lack of accurate heterogeneous thrombus analogues [14,27,28], [29]. To address these shortcomings, we propose developing digital twins of thrombi —biology driven, in silico representations at both macro-scale (~mm) and micro-scale (~μm) — enabling virtual testing and characterization. This approach aligns with a broader shift in biomedical research, where advanced computational models facilitate comprehensive analyses through virtual clinical trials [27,30].

Various computational tools have been developed to model fiber network microstructures, including those of fibrin and collagen. There are image-based models [31] and simple triangulation models based on Delauney [32], or Voronoi networks [33]. Simplistic triangulation models of fibrin network structures such as [32], cannot accurately replicate key characteristics of fibrin network such as connectivity, fiber length, and directionality. More advanced tools utilize confocal microscope image stacks to create 3D network reconstructions [31,34]. However, these approaches are constrained not only by the limited imaging depth of optical microscopy but also by practical challenges common to experimental studies, such as non-representative thrombus analogues and the difficulty of obtaining patient thrombi in sufficient quantities. More recent generative pipelines have been developed to generate realistic fiber network topology [33,35,36]. However, these models lack the integration of red blood cells (RBCs), platelets, and the heterogeneous spatial organization characteristic of real thrombi.

Thrombi also exhibit a wide range of size and macroscopic geometries, influenced by vessel anatomy, their origin, and patient-specific factors. Thrombus dimensions can range from hundreds of microns to several centimeters [37,38], and this variability is reflected in their photoacoustic response [39].

In this work, we develop REFINE, a framework for generating in silico thrombi replicating key statistical features of the microstructure and heterogeneity, incorporating fibrin network, RBCs, and platelets. REFINE can generate unique structures conforming to literature extensive microstructure descriptions of thrombi (e.g. fibrin fibers length and connectivity, heterogeneous distribution of RBCs and platelets, and porosity). REFINE distinguishes itself from previous models by its topology-driven recursive optimization, which allows precise control over fiber network properties. This makes it the first framework capable of integrating structural accuracy with computational efficiency for both photoacoustic and future biomechanical simulations. These in silico thrombi are then simulated by a novel multiscale photoacoustic simulation platform to complete our digital twin model of thrombus photoacoustic imaging. We report here the first multiscale photoacoustic simulation platform and demonstrate its utility in filling the gap for accurate signals' reproduction for heterogeneous micro structured tissue types [40,41].

The paper is structured as follows. First, the thrombus microstructure modelling pipeline (REFINE) for matching the realistic parameter distribution of fibrin network parameters, RBCs and platelets is introduced. Then, the multiscale photoacoustic simulation platform will be described. Next, two simulation scenarios are presented to demonstrate the advantages of multiscale simulation in thrombus characterization compared to conventional methods, showing strong agreement with experimentally reported results. Additionally, we create two thrombus phantoms and record the photoacoustic responses in the lab for empirically testing our framework. Finally, we discuss the potential of the proposed in silico thrombus modelling and simulation platform for advancing diagnostic photoacoustic imaging.

## 2. Methods

### 2.1 Recursive Iterative Fibrin Network Emulation (REFINE)

Fibrin network is a critical determinant of thrombus integrity. Fibrin monomers polymerize and crosslink to form a mesh that traps blood cells and reinforces the structural stability of the thrombus. This process results in a complex microstructure with variations in density, branching, fiber dimensions, and orientations. Ultimately, these factors influence the thrombus mechanical behavior [42]. Therefore, implementing an algorithm capable of tailoring these parameters while generating a representative thrombus model is crucial for accurate in silico modeling.

To describe the fibrin network within a thrombus, several key terms must be defined. We refer to any spatial location (x, y, z) where a fibrin strand terminates or intersects as a *node*. Connections between nodes are referred to as *fibrin fibers*. *Connectivity* is defined as the number of fibers per node, and each fiber can vary in length (*fiber length*). Additionally, the *direction cosine distribution* describes the cosine of angles between fibers at a given node. These parameters are typically characterized by probability distribution functions (PDFs) derived from experimental confocal imaging data [42].

The REFINE algorithm initializes the fibrin network as a random structure and iteratively optimizes it to match the target parameter distributions described above. The discrepancy between the current and target node distributions is quantified using the unitless Jensen–Shannon divergence (JSD) metric. During optimization, forces are applied to nodes—adjusting fiber lengths and orientation—to progressively align the network with the target PDFs. The optimization process is guided by parameter-specific weights. In this work, we assign equal weight to all parameters.

The REFINE algorithm accepts a total number of RBC inclusions, overall volume of inclusions and total fibrin concentration and ratio of platelet crosslinks (connectivity = 4) over the total number of cross links as inputs.

The process begins by generating spherical, non-intersecting volumes with random radii within the target thrombus domain to allocate space for RBC inclusions (step 1, Fig. 1). The initial positioning and radii of the inclusion spaces determine the overall shape of the clot. Thus, there are no limitations on thrombus shape and volume, and the generated thrombus microstructure can conform to any arbitrary geometry.

In the next step (step 2, Fig. 1), nodes are seeded around these inclusions, ensuring they do not fall inside the inclusion volumes. To achieve this, we randomly position nodes according to 3D multivariate normal probability distribution functions with means located at the sphere centers, as defined

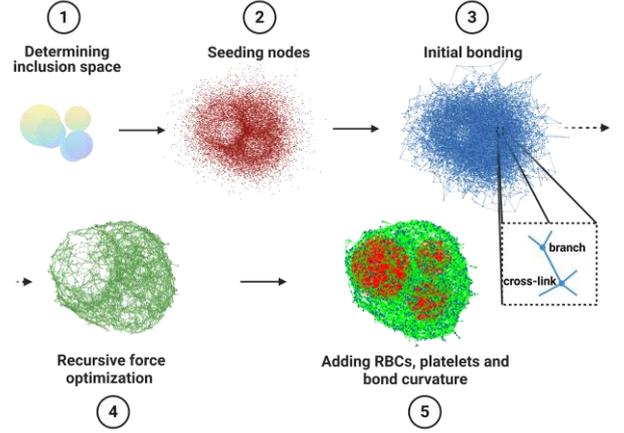

**Figure 1.** The stepwise process to generate thrombus microstructure using the REFINE algorithm

in Equation 1. Nodes that fall inside the spheres are then filtered out.

$$PDF(p_n, \mu, \Sigma) = \frac{1}{\sqrt{|\Sigma|(2\pi)^3}} e^{(-\frac{1}{2}(p_n - \mu)\Sigma^{-1}(p_n - \mu)')} \quad (1)$$

Here, $p_n$ and $\mu$ are 1×3 vectors representing the coordinates of the target node and the center of the current spherical inclusion, respectively. $\Sigma$ is a 3×3 symmetric, positive-definite matrix, where the diagonal elements define the variances of each variable, and the off-diagonal elements represent the covariances between variables. In this case, we assume an isotropic distribution with no covariance. Additionally, the variances are chosen to ensure enough points are generated in the spaces between inclusions.

This step also involves refinement of nodes. In realistic thrombus structures, the density of the fibrin network is not spatially uniform, with denser regions typically forming between inclusions due to outward growth forces originating from RBC-rich inclusions. To account for this effect, a displacement function is applied to the surrounding nodes, pushing them outward relative to their distance from the inclusion centers. The resulting displacement vector $\vec{r}$ for each node can be calculated as:

$$\vec{r} = w(d)\hat{d} , \quad \hat{d} = \frac{C_s - p_n}{\|C_s - p_n\|} , \quad d = \|C_s - p_n\|$$

$$w(d) = \begin{cases} a\left(1 - \dfrac{d - r_s}{r_s}\right) & r_s \leq d \leq 2r_s \\ 0 & d > r_s \end{cases} \quad (2)$$

where, w(d) is a distance-dependent weight function, d represents the distance between each node and the inclusion center ($C_s$), and $\hat{d}$ is the unit vector pointing from the inclusion center toward the node $p_n$. The inclusion radius is denoted by $r_s$, while $a$ is an arbitrary coefficient that defines the magnitude of the displacement.

With the initial node positions determined, the initial fibers are formed (step 3, Fig.1). This is done using the nearest neighbor law. First, the connectivity number for each node is randomly generated based on an experimentally measured PDF reported in [42]. The established law regarding fibrin networks reports a mean node connectivity Z between '3' (branching) and '4' (crosslinking). According to the quantitative estimations of [42] for a fibrin network with mean connectivity of 'Z', the connectivity distribution PDF $N(p)$ is a shifted geometric distribution expressed by:

$$N(p) = q(1-q)^{p-3} , \quad q = \frac{1}{Z-2} \qquad (3)$$

where p is the connectivity (≥3) and Z is the mean connectivity through the whole network, set using the input fibrin concentration according to interpolated values from [42].

The next step involves applying the recursive force optimization method (step 4, Fig.1), where we minimize JSD metric for the fibrin length distribution and fibrin orientation (direction cosine distribution). The experimental fiber length distribution in a fibrin network can be approximated as a lognormal distribution, as characterized in [42]. The average fiber length is linearly extrapolated as a function of fibrin concentration based on experimental data. According to [42], the measured average fiber lengths are approximately 4.87μm at a fibrin concentration of 0.4gr/L and 2.99 μm at 1.6gr/L. The direction cosine distribution is also an experimentally measured quantity and can be either isotropic or anisotropic.

The JSD metric is computed as shown in (4) to optimize the predetermined target distributions [43].

$$JSD(P\|Q) = \frac{1}{2}D_{KL}(P\|M) + \frac{1}{2}D_{KL}(Q\|M)$$
$$M = \frac{1}{2}(P+Q) \qquad (4)$$
$$D_{KL}(P\|Q) = \sum_{x \in \mathbb{Z}} P(x)\frac{P(x)}{Q(x)}$$

where '$D_{KL}$' is the Kullback-Leibler divergence [43] and P, and Q are the two probability distribution functions (achieved vs. targeted) to be compared. The JSD provides a symmetric similarity metric between 0, when the two PDFs are identical, and 1, when the two PDFs are completely dissimilar. We compute the JSD at each iteration of the recursive force optimization loop for all the parameter distributions with respect to the desired target distributions. If the combined divergence metric (weighted sum of all parameter's JSD metrics) is less than the optimization target error the points need to be reconfigured/moved towards the desired configurations. If the current average fiber length is larger than the target averages, a displacement vector (relaxation force) for each pair of nodes connected with a fiber is calculated (Fig.2). This relaxation force is expressed as:

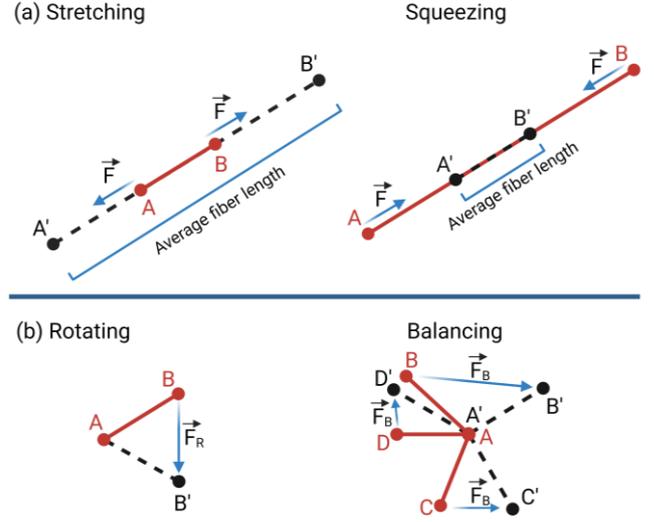

**Figure 2.** Diagram of force and displacement direction. (a) optimizing the fiber length (squeezing and stretching), (b) optimizing the angle (rotation and balancing)

$$|\vec{F}| = k_{Fibrin}|d_0 - d_{average}| \qquad (5)$$

following Hooke's law, with the target average fiber length '$d_{average}$' representing the equilibrium fiber length. '$d_0$' is the current value of the fiber length for a certain fiber (e.g. AB, Fig.2a) and '$k_{Fibrin}$' is spring constant assuming a simplified spring model for fibers. This can be derived from Young's modulus of fibrin varying between 14.5 and 23 MPa [34] and its given length and thickness [44]. Amplitude of rotation ($F_R$) and balancing ($F_B$) forces (Fig.2b) is derived by calculating the required displacement vector based on the desired directionality vector or required displacement vector for balancing each node.

To ensure stability and improve convergence speed, the force amplitudes are dynamically adjusted based on the JSD metric's rate of change. To prevent trapping in local minima, random perturbations are introduced to the node positions. Outlier nodes and fibrin fibers that intersect RBC inclusion volume are then removed.

Finally, in step 5, fibrin fibers are assigned curvature using 3-point Bezier curves. Ideally, the curvature distribution should reflect experimental measurements of fibrin fibers in the thrombi. However, due to the lack of available data, we assume a random uniform distribution.

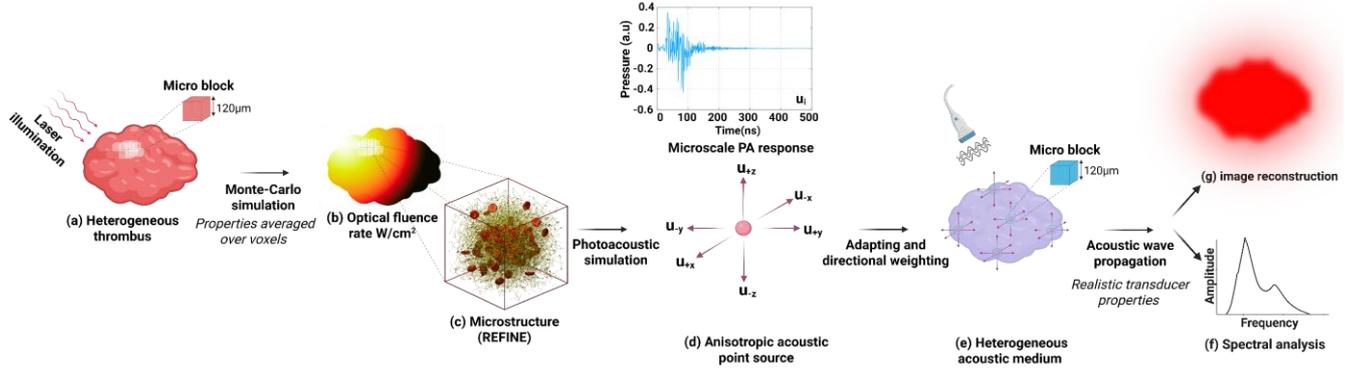

**Figure 3.** Multiscale photoacoustic simulation approach

After the fibrin network is created, platelets and RBCs are added to complete the thrombus structure. Platelets attach to fibrin fibers via their filopodia; however, due to their small size (2 - 3 μm diameter) [45] and the resolution of our simulations (0.5 μm), platelets are approximated as spheres. We assume platelets to be randomly positioned on a portion of the crosslink nodes.

RBCs are simulated as biconcave disks with diameter of 7-8 μm, using the surface equations described in [46]. The total number of RBCs is calculated based on the radii and number of inclusion spaces which are given as input values. RBCs are then individually rotated, translated, and homogenously compressed (up to 30%) to fill up and match the predefined inclusion volumes.

### 2.2 Multiscale photoacoustic simulation approach

We introduce a multiscale photoacoustic simulation pipeline (Fig.3) that captures the macro- and microscale features of thrombus, while mitigating the computational cost of high-resolution simulations over large domains. This is achieved by integrating microscale photoacoustic responses into a macroscale simulation framework. We discretize the heterogeneous 3D thrombus volume into microstructure blocks. The dimensions of these blocks are determined by the thrombus heterogeneity and the computational constraints related to overall thrombus size. Additionally, the blocks cannot be smaller than the photoacoustic thermal and stress confinement scales for a given laser pulse width [47].

We characterize the thrombus by its porosity and RBC composition at both macro (global) and micro (local) scales (Equ. 6).

$$RBC\ composition = \frac{v_{RBC}}{v_{RBC}+v_{Fibrin}+v_{Platelet}} \quad (6)$$

$$porosity = 1 - (v_{RBC} + v_{Fibrin} + v_{Platelet})$$

where $v_{RBC}$, $v_{Fibrin}$, and $v_{Platelet}$ are volume fractions of RBCs, fibrin fibers, and platelets, respectively. Size, shape and heterogeneity of thrombus determine the volume fraction of constituents over its volume. At each point we randomly generate a matching microstructure block exhibiting the local porosity and RBC composition.

The first step of the multiscale simulation (Fig.3a and Fig.3b) involves computing the macroscale optical energy deposition using ValoMC, an open-source Monte-Carlo photon packet simulator [48]. To perform this simulation, the optical absorption coefficient ($\mu_a$), optical scattering coefficient ($\mu_s$), refractive index, and optical anisotropy factor must be specified at each mesh coordinate. The mesh is tetrahedral shaped (mesh size = 120μm). These optical parameters are estimated from the volume fraction of RBCs, fibrin, and platelets for each block, knowing their individual bulk optical properties at the desired wavelength (532 nm). We assume that the empty spaces inside the thrombus are filled with blood plasma. The Monte-Carlo simulation produces an irradiated optical energy distribution, represented as the optical fluence rate $\Phi(x,y,z)$ (W/cm$^2$), throughout the entire thrombus.

In the next step (Fig.3c), we assign microstructures to each block using our thrombus generation algorithm (REFINE). We create small thrombus (5nL in volume) based on the known porosity and RBC composition of the given block and extract a cubic sub-volume of 120 μm × 120 μm × 120 μm from it.

Subsequently, microscale photoacoustic simulation is performed. To accurately capture the fine microstructural details while considering computational limitations, we use a mesh size of 0.5μm in these simulations. The initial photoacoustic pressure ($p_0$) profile inside the microstructure is then calculated by (Equ. 7).

$$p_0 = \Gamma \mu_a \Phi \quad (7)$$

where $\Gamma$ is the Grüneisen parameter [47], $\mu_a$ is absorption coefficient at each mesh grid point.

The microscale acoustic simulation of each block is done utilizing k-Wave time-domain acoustic simulator [49]. We capture the acoustic responses at the boundaries of each block by imposing a perfectly matched layer (PML) boundary condition. By averaging the acoustic response over each of the six facets, we store the result as a particle velocity point source with six directional components, each corresponding to a facet of the microstructure block (Fig. 3d).

While this approach preserves directional information from the microscale structure, macroscale photoacoustic response is also anisotropic particularly in heterogeneous or structured media [19].

The directionality of the photoacoustic response is dictated by the initial particle velocity $u_0(r)$. Assuming $u(r,t)$ as the particle velocity vector we can invoke conservation of momentum:

$$\rho \frac{\partial u(r,t)}{\partial t} = -\nabla p(r,t) \qquad (8)$$

where $\rho$ is density. The pressure wave generated in a photoacoustic process can be derived from photoacoustic wave equation [19]:

$$\nabla^2 p(r,t) - \frac{1}{c^2}\frac{\partial^2 p(r,t)}{\partial t^2} = -\frac{\beta}{C_p}\frac{\partial H(r,t)}{\partial t} \qquad (9)$$

Here $p$ is the pressure, $\beta$ is the thermal coefficient of volume expansion, $C_p$ is the specific heat constant at constant pressure and $H(r,t)$ represents the heat distribution. Considering stress and heat confinement, $H(r,t)$ is modeled as an instantaneous delta pulse: $H(r,t) = H(r)\delta(t)$. With this simplification, the initial pressure $p_0 = \Gamma H(r)$ where $\Gamma$ is the Grüneisen parameter. Integrating (8) immediately after $t=0$, and assuming the pressure change occurs over a short time $\Delta t$, we find:

$$u_0(r) \approx -\frac{1}{\rho}\nabla p_0(r) = \frac{\Gamma}{\rho}\nabla H(r) \qquad (10)$$

According to (10), the initial particle velocity is directly proportional to the absorbed energy gradient, which is most likely anisotropic considering the illumination profile.

To better reflect this physical behavior, we introduce directional weighting to the six velocity components based on the gradient of absorbed energy. This approach accounts for the anisotropy in initial wave propagation that arises from both asymmetric illumination and heterogeneous microstructure—features that are typically underrepresented in standard simulations using only scalar initial pressure. The normalized local gradient of absorbed energy computed from the Monte Carlo field $H(r)$ is calculated at the center of each voxel $r_c$ as

$$\vec{w} = \frac{\nabla H(r_c)}{|H(r_c)|} \qquad (11)$$

The velocity point source directional components are then scaled as the dot product of the weight vector and point source vector in the normal of the cubic voxel direction (Fig. 3d):

$$q_n(t) = \vec{w}\cdot\vec{u}_n(t) \qquad (12)$$

Where $\vec{u}_n(t) = u_n(t)\hat{e}$ and $n \in \{\pm x, \pm y, \pm z\}$ and $\hat{e} \in \{\pm\hat{i}, \pm\hat{j}, \pm\hat{k}\}$ the unit vectors.

The final simulation step (Fig. 3e) incorporates these directionally weighted point sources in a fully acoustic simulation to find the acoustic response at the detector position. In this step, we employ the averaged acoustic parameters (e.g., density and speed of sound) over microstructure blocks into the thrombus macroscale model and place corresponding anisotropic acoustic point source at the center of each block.

This process enables the photoacoustic simulation of the whole thrombus with significantly reduced computational demands compared to a fully resolved microscale simulation at 0.5 μm resolution. The simulated radiofrequency (RF) signals are also band pass filtered to reflect the empirical bandwidth limitations.

### 2.3 Simulation setup

To demonstrate the significant advantages of our multiscale framework over conventional simulations, we create virtual thrombus samples with differing microstructure and composition. This simulation scenario aligns with previously reported experiments [26,39].

To introduce physiologically relevant microstructure and composition differences, we generate 12 homogeneous cylindrical samples (thickness = 1 mm, diameter = 6 mm), divided into four groups designed to span a broad range of thrombus stiffness. The groups are defined based on RBC content and porosity as follows: RBC composition ≥ 95% and porosity ≤ 30% (Group 1); RBC composition ≥ 90% and porosity ~ 50% (Group 2); RBC composition ~ 70% and porosity ~ 80% (Group 3); RBC composition ≤ 50% and porosity ≥ 95% (Group 4). These compositional differences correspond to a wide range of mechanical stiffness, as previously reported in [14]. Each sample contains 50 unique microstructure blocks (generated by REFINE), randomly distributed throughout the volume to represent local microstructure variations. These blocks and the corresponding microscale photoacoustic responses were generated prior to simulations. Each block is implemented in MATLAB on a PC with an Intel (R) Xeon(R) CPU @ 3.70GHz, 10 Core(s), 64GB of RAM and NVIDIA RTX A4000 GPU. Also, acoustic responses of each block are simulated using k-Wave with GPU acceleration. A full thrombus of the above-mentioned size consists of 17649 blocks.

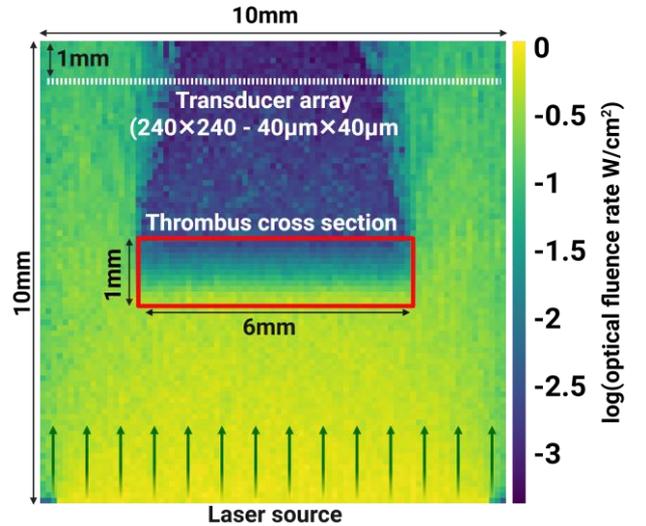

**Figure 4.** Simulation setup schematics (2D cross-section). The colors indicate the logscale normalized fluence rate (logscale).

**Table 1.** Optical ($\lambda = 532\ nm$) and acoustic parameters of the thrombus components used in our simulations. Values used in this work are reported in bold. Sources used are [31,51–56]

| Component | $\mu_a$ (mm$^{-1}$) | $\mu_s$ (mm$^{-1}$) | Speed of Sound (m/s) | Density (kg/m$^3$) | Refractive Index |
|---|---|---|---|---|---|
| RBCs | 23.0 | 70 – 100 | 1570 | 1125 | 1.38 – 1.41**(1.40)** |
| Platelets | 0.2 – 0.8 **(0.5)** | 0.05 – 0.2 **(0.125)** | 1540 | 1060 | 1.35 – 1.40 **(1.38)** |
| Fibrin | ≪ 0.005 **(0.005)** | 0.06 | 1555 – 1580 **(1567.5)** | 1080 | 1.53 – 1.62 **(1.58)** |
| Plasma | 0.02 – 0.1**(0.49)** | 0.05 – 0.2 **(0.125)** | 1510 – 1540 **(1525)** | 1025 – 1030 **(1027.5)** | 1.34 – 1.35 **(1.35)** |

Our setup (Fig.4) uses a wide-beam laser illumination from the bottom, with a cosine beam profile (diameter = 10 mm) in 532 nm and pulse width of 5 ns. It also contains a 2D array of transducers on the top (240×240 array of 40×40 μm sensors) and the sample in the middle. This transducer array is also bandwidth limited to have more relevance to the experimental limitations. We use a flat-top frequency response (center frequency : 5MHz – 80% bandwidth) in our simulations, as this range is commonly used in both experimental research setups and clinical ultrasound imaging systems [50].

A mesh conversion step is needed from the optical simulation tetrahedral shape to the rectangular shape constrained by k-Wave simulation software. We choose the mesh size in the acoustic simulations to be 40 μm to cover the transducer frequency bandwidth reliably (maximum frequency ~ 19MHz) [49]. The simulation time step is also set to 0.5ns to be consistent with microscale responses. The optical and acoustic parameters for RBCs, platelets, Fibrin and blood plasma in our simulation have been extracted from experimental reports in [31,51–56] and summarized in Table 1.

To have a comparison reference, we also simulate the same samples using a typical but simplistic photoacoustic simulation. We assume all simulation parameters to be the same. However, the simplistic simulation directly uses optical simulation output multiplied by Grüneisen parameter as initial pressure input without taking the microstructural responses into account.

Moreover, to validate the spectral trends and image features observed we perform simple experimental measurements on agarose tissue mimicking phantoms embedded with polystyrene, black-dyed microspheres (6 μm diameter; Polybead®, Polysciences Inc., Cat. No. 17135). The photoacoustic signals at 532 nm for phantoms of 40% and 15% bead concentrations are measured in a transverse broad illumination photoacoustic setup using a linear ultrasound array (L11-5v, 300 μm pitch, Verasonics inc., Kirkland, USA). The transducer center frequency matches that used for the simulated data (5 MHz), albeit with a differing impulse response. The pixel size is defined by the transducer and is thus larger for the experimental images.

To analyze the simulation outputs and assess performance of our multiscale simulation approach, we compare beamformed images of simplistic simulations, by calculating the normalized variance of pixel intensities, defined as the variance divided by the mean intensity within a 1 mm × 1 mm region of interest and calculating representative spectra of the samples (utilizing Principal Component Analysis (PCA), [57]). Furthermore, spectral sharpness defined as peak frequency over 3 dB bandwidth are compared for the different numerical and lab experiments.

## 3. Results

### 3.1 Virtual thrombus generated by REFINE

The REFINE algorithm can create a wide range of thrombus microstructures with varying heterogeneity, porosity, and composition, with different thrombus shapes. Fig.5 shows an example of a thrombus generated by our REFINE algorithm. In this example, we have 10 inclusion spaces filled with RBCs and a fibrin network with fibrin concentration of 1.6g/L is created around them. The resulting thrombus has a volume of 6.55nL. The thrombus model is generated under the assumption of isotropic fibrin distribution, using target structural metrics, such as connectivity, fiber length, and orientation distribution derived from literature [42].

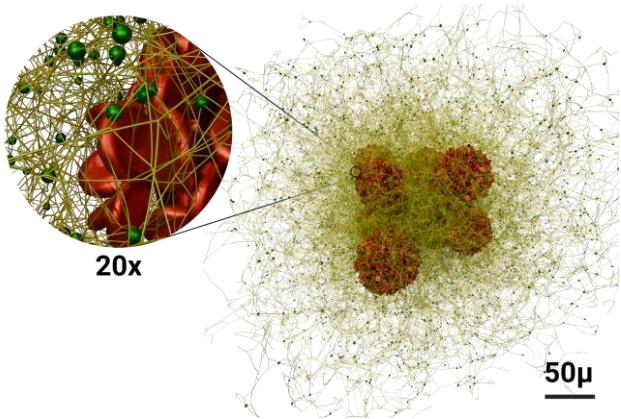

**Figure 5**. Example of a thrombus sample generated with our REFINE algorithm. RBCs are shown as red disks, platelets as green spheres, and fibrin fibers as yellow lines.

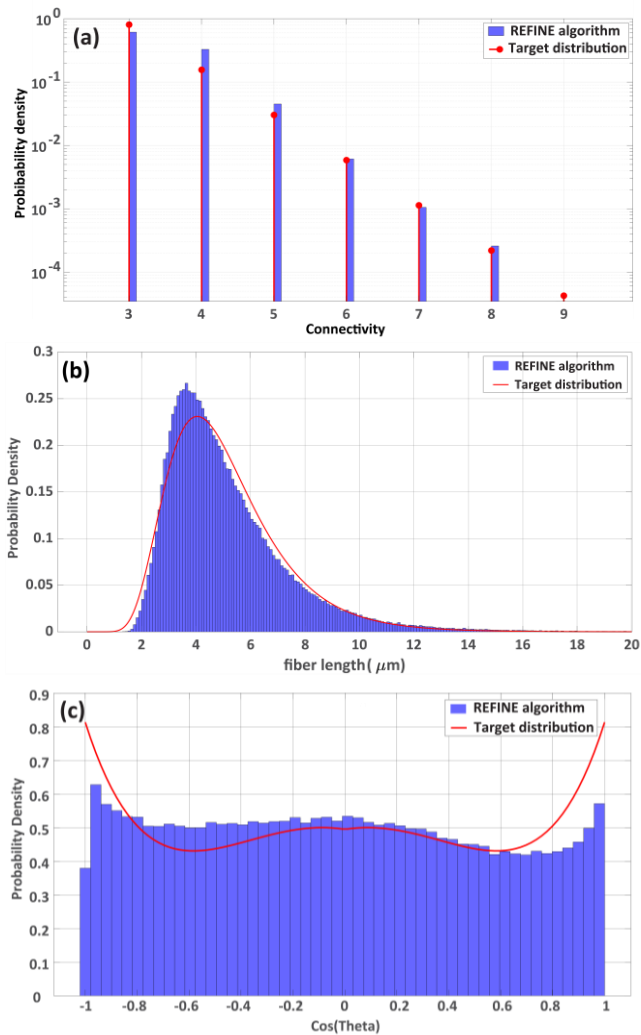

**Figure 6**. Comparison of parameter probability distributions in the generated thrombus (REFINE algorithm) with the corresponding target distribution from experimental measurements [42]. a. Connectivity distribution b. Fiber length distribution. c. Direction cosine

Fig.6a,b compares the calculated parameter distributions with the fitted distributions extracted from confocal microscopy images. Fig.6c also shows the calculated direction cosine distribution for our generated thrombi which is following the same trend as [42]. The target distributions are well achieved with a maximum error of 12%, 5%, and 11%, respectively.

To assess robustness to random initialization, the REFINE algorithm was executed 20 times using identical input parameters. Convergence times varied across runs, with most optimizations completing within 0.18–0.48 seconds (mean $0.23 \pm 0.07$ seconds). Despite this variability in runtime, the final microstructural metrics were highly consistent and porosity and composition values across runs converged to $0.87 \pm 0.004$ and $0.60 \pm 0.02$ (mean ± SD) respectively, indicating reliable attainment of the target distribution regardless of initial seeding.

### 3.2 Multiscale simulation

We compare the multiscale and typical simulation approaches on the 12 generated thrombi. The reconstructed images for both methods are depicted in Fig.7a,b for a sample thrombus (RBC composition ~ 70% - Porosity ~ 80%) respectively [42]. The typical photoacoustic simulation shows a solid-like appearance while the multiscale approach bears more information and has speckles originating from its microstructure. The RBC clusters, which are the primary absorbers, exhibit random variations in size, shape, and orientation. Their 3D spatial distribution is governed by the density and heterogeneity of the surrounding fibrin network. Fig.8a, b illustrates two sample microstructures generated by REFINE, highlighting distinct structural differences. These microstructural variations influence the time-domain photoacoustic responses and introduce anisotropy. While conventional photoacoustic simulations of thrombi often neglect such microstructural effects—due to computational constraints and the assumption that they are negligible relative to the acoustic wavelength—the cumulative impact of these small variations can significantly alter the photoacoustic response, particularly in the spectral domain. Consistent with the multiscale simulation, the tissue mimicking phantoms (Fig. 7c) also has a speckle like appearance, confirming the need for such a simulation framework. The typical simulation (Fig. 7a) showed a low normalized variance of 0.02, whereas the multiscale simulation (Fig. 7b) and the phantom sample (Fig. 7c) exhibited comparable values of 0.78 and 1.12, respectively, confirming good agreement between simulation and experiment.

A PCA-based general representative spectrum is extracted

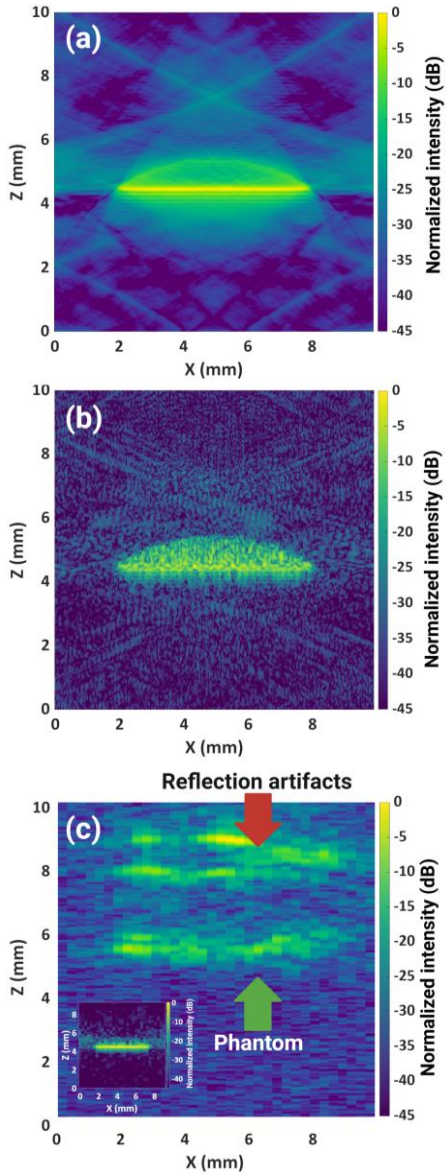

**Figure 7.** Sample 2D cross section of the reconstructed PA image. The simulated thrombi show ~70% RBC composition and ~80% porosity a. Typical simulation b. Multiscale simulation (both images are log-scale) c. Beamformed image of the phantom sample (15% microbead) showing similar speckles as multiscale simulation (Green arrow points to the phantom and red arrow points to reflection artifacts). Inset is the beamformed data from the multiscale simulation in (b) processed to match the (L11-5v) transducer bandwidth and pitch

from the 240 × 240 signals recorded. In Fig.9a,b normalized to the maximum PCA spectrum for our multiscale simulation and the typical simulation approaches are illustrated for the four sample groups. In Fig. 9c we observe similar trends for the simplistic tissue-mimicking phantom experiment. To validate that our findings align with previously reported experimental patterns we quantify spectral differences between the groups by calculating a metric expressed as peak frequency over the bandwidth for all samples using our multiscale simulation approach in Fig.10.

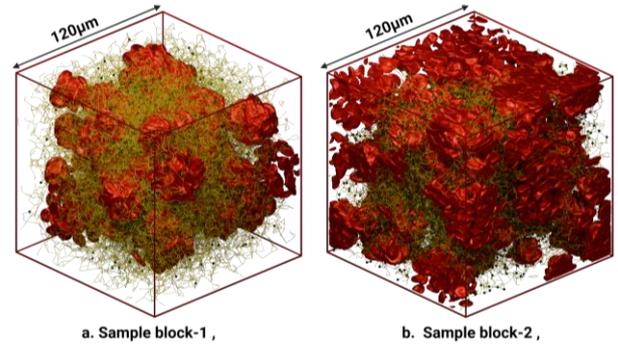

**Figure 8.** Sample microstructure blocks a. ~70% RBC composition and ~80% porosity. b. ~90% RBC composition and ~50% porosity. RBCs are shown in red, Fibrin in yellow and platelets in green.

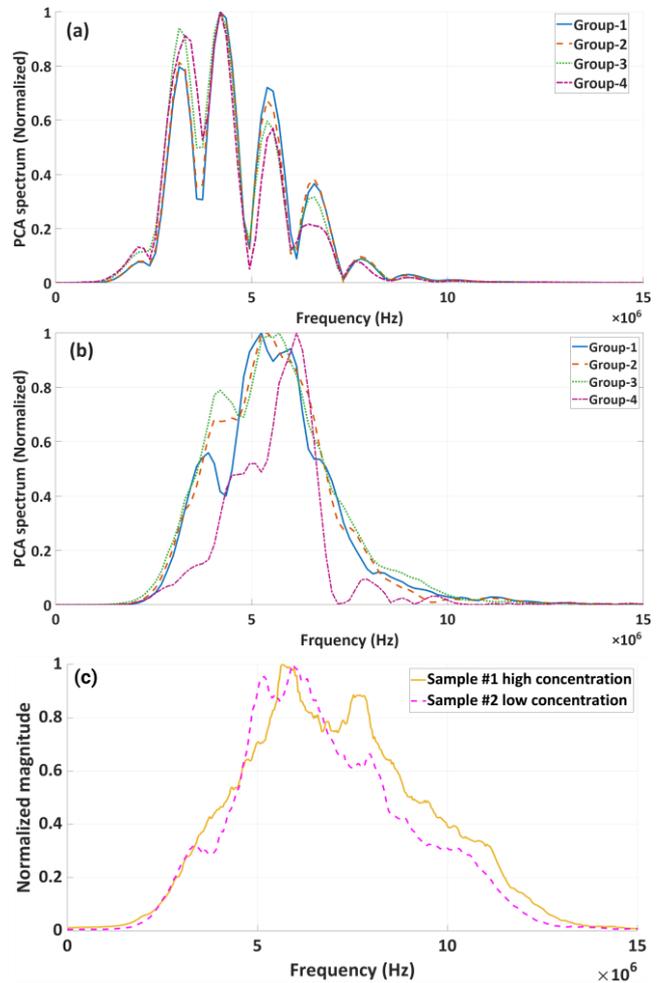

**Figure 9.** PCA representative spectrum for (a) Typical PA simulation for sample groups showing similar behavior. (b) Multiscale simulation approach for the same sample groups showing significant spectral variations (one variation per group). (c) Measured PCA spectral data for the two phantoms with low and high bead concentration.

According to Fig.10 this metric clearly correlates with RBC composition and porosity (thrombus stiffness) as experimentally verified in [26,39], with the stiffness decreasing progressively from Group 1 to Group 4, based on [14]. Ultimately, the typical simulation approach fails to provide enough insight regarding spectral differences. However, we observe notable differences in the photoacoustic spectral behavior of the samples, particularly in key features such as bandwidth and spectral power distribution using our multi-scale approach. Stiffer thrombi consistently exhibit broader bandwidths, with spectra shifted toward lower frequencies. In contrast, highly porous samples produce a narrower spectrum shifted toward higher frequencies. The presence of lower frequencies and wider bandwidths likely relates to the higher density and semi-solid structure of the stiffer samples. In contrast, increased porosity and lower RBC content appear to produce higher-frequency components, likely due to isolated clusters of trapped RBCs. A similar trend relating to the thrombus stiffness [39] and RBC occupational percentage [26], is reported in previous experimental studies and also observed in the tissue mimicking phantom experiment performed in this study. The spectral sharpness metric was found to be 0.827 for the high-concentration (40%) sample and 0.904 for the low-concentration (15%) sample, confirming the trend observed in the multiscale simulations.

Finally, we report the simulation runtime of the multiscale simulation of each sample. The whole simulation run time for multiscale scenario from assembling microstructure blocks to form the thrombus sample, microscale photoacoustic simulation, running the Monte-Carlo optical simulation and final k-Wave acoustic simulation is about 250 minutes.

## 4. Discussion

### 4.1 REFINE Framework

REFINE has demonstrated strong potential in replicating key microstructural traits of real thrombi. It generates thrombi with controllable heterogeneity, porosity, and cellular composition by combining topology-driven network initialization, iterative relaxation guided by statistical distributions, incorporation of inclusions, and assembly into macroblocks. Although currently tailored to thrombus, where fibrin networks and cellular inclusions dominate, the core methodological steps are not thrombus specific. For other heterogeneous tissues such as tumors or fibrotic lesions, appropriate target distributions (e.g., collagen fiber orientation, anisotropy, cellular cluster size, necrotic voids) could be defined, after which the optimization algorithm would adjust the microscale topology until the synthetic network matches these descriptors. The downstream voxelization and photoacoustic simulation steps remain unchanged, making REFINE extensible to a broad range of biological tissues.

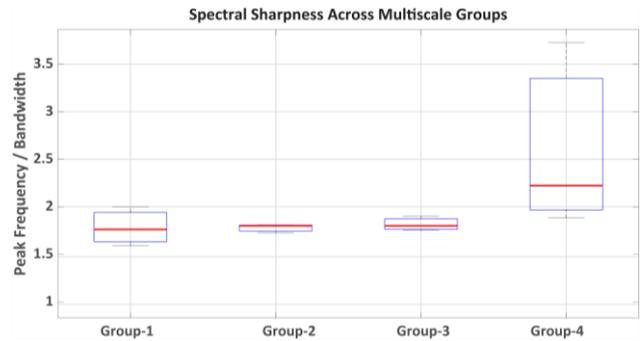

**Figure 10**. Peak frequency/bandwidth as a microstructure dependent metric based on multiscale simulation approach.

In the current version of REFINE, we adopt some pragmatic simplifications that establish a foundation for subsequent refinement. Fibrin fibers are assigned uniform curvature distributions in the absence of detailed experimental measurements of fiber waviness and contracted networks descriptors. The parameters distribution is aligned with the non-contracted fibrin networks [42]. Platelets are modeled as spheres, neglecting activation-dependent morphology; and constant optical anisotropy factor is assumed for each microstructure block in the photon-packet Monte Carlo simulation. These simplifications enable validation of the framework while none are fundamental barriers: curvature distributions, realistic platelet geometries, optical anisotropy, and more efficient implementations can be integrated as data and resources allow. To further enhance its realism and utility, future developments could incorporate thrombus contraction, thrombus aging, and patient-specific variations. Contraction may be represented by adjusting fibrin network connectivity and density, while aging effects could be modeled through changes in cross-link density, RBC shape, and optical absorption coefficient changes. Patient-specific variation could be incorporated by tailoring the algorithm inputs to target distributions extracted from patient-derived histology or imaging data.

### 4.2 Multiscale photoacoustic simulation framework

By generating large virtual cohorts of thrombi with biologically relevant microstructural variability, the framework provides the means to statistically link photoacoustic spectral biomarkers to thrombus composition. This capability not only complements experimental efforts but also establishes a scalable platform for training machine-learning models aimed at thrombus characterization.

In this study, a wide-beam illumination geometry has been demonstrated. However, catheter-based geometries can be easily incorporated in future studies, by adjusting the illumination source and receive geometry and bandwidth. The developed framework can also be used to investigate the effect of randomly distributed RBCs in the blood surrounding the thrombus.

Generally, the spectral features of the photoacoustic signal are strongly influenced by the underlying microstructure of the sample. Phantom experiments using agarose gels embedded with polystyrene microbeads were conducted to validate the framework under controlled conditions. By varying microbead concentrations, we were able to test whether absorber concentration (sample microstructure) leads to predictable spectral shifts and speckle formation for the acoustic bandwidth and sample dimensions simulated. The phantoms demonstrated spectral trends consistent with the multiscale simulations, and the speckle patterns are present in the photoacoustic images. However, measurements on more realistic tissue-mimicking phantoms which include fibrin networks, or on well-characterized thrombus analogues are needed to provide a one-on-one reliable comparison of this framework generated images, which is the object of future studies.

Photoacoustic simulation with high microstructure accuracy requires immense computational resources and extended runtimes—even for relatively small volumes—to accurately reproduce the photoacoustic response. This challenges the practical utility of computational modeling, particularly when the goal is to generate a large number of clinically representative samples for in silico analysis. In contrast, our multiscale simulation approach captures the critical correlations between microstructure and spectral behavior more effectively, while reducing computational time to ~250 minutes per sample. While improved, the simulation times remain long, limiting capability to simulate large sample numbers. Each multiscale simulation requires generating microstructures (~2 minutes per each block), calculating their microscale responses (~30 seconds per variation), assembling the macroblocks to create the whole macroscale shape (~60 minutes), Monte-Carlo optical simulation (~1 min) and final acoustic simulation (~60 minutes). In future work, simulation run times can be reduced through more efficient implementation of the REFINE algorithm, including lowering memory requirements, parallelizing computational steps, and incorporating GPU acceleration for large-scale vector calculations. We also plan to re-implement the microblock assembly process using optimized GPU-based matrix operations to further decrease both memory usage and assembly time. Ultimately, developing a unified in-house implementation of the entire pipeline, rather than relying on multiple third-party libraries, could substantially shorten overall simulation times.

The overarching goal of this in silico pipeline is to overcome the constraints of experimental and clinical settings by generating large-scale datasets for investigating thrombus microstructure and biomechanics—particularly through machine learning and other spectral domain analysis techniques. Despite promising results, the multiscale framework and in silico thrombus generation method require further validation using ground-truth microscopic imaging. Nonetheless, the approach effectively balances biological accuracy and efficiency, enabling large-scale in silico studies that would otherwise be infeasible

## 5. Conclusion

In this work, we introduced REFINE, a multiscale computational framework for in-silico thrombus generation and photoacoustic simulation. By combining biologically relevant thrombus generation with a multiscale photoacoustic simulation pipeline, we demonstrated the critical role of such method to capture thrombus microstructure information from photoacoustic spectral responses at different scales. We show that our multiscale approach reproduces key spectral trends consistent with experimental observations, which are otherwise missed using classical simulation approaches.

REFINE enables the generation of large-scale, biologically accurate virtual thrombus datasets, supporting the exploration of microstructure-related spectral biomarkers. Looking ahead, integrating this digital twin model with machine learning could allow real-time thrombus characterization during intravascular photoacoustic-guided interventions, paving the way for personalized treatment strategies. Moreover, REFINE's recursive optimization and topology-preserving design make it well-suited for extension beyond imaging. In summary, REFINE offers a unique blend of geometric flexibility, biological realism, and parameter-driven control. Unlike previous generation techniques, it supports customizable topology and spatial heterogeneity, laying a solid foundation for multiphysics modeling, including structural and mechanical simulations.

While promising, further improvements are needed, including the integration of ground-truth data, expansion to larger experimental datasets, and incorporation of mechanical models. Nonetheless, this study establishes a key step toward the use of in silico thrombus modeling to advance diagnostic imaging and treatment planning in thrombosis-related diseases.

## Acknowledgements

This work was partly supported by the Dutch Research Council (NWO) under the grant OCENW.XS24.2.210 awarded to SIR and the Delft University of Technology Cohesion Grant awarded to SIR and BF.

## Data availability

To support reproducibility and broader use, the REFINE algorithm and code are openly available at: https://github.com/hghodsi7980/ThromboGen.